# An experimental study of the detailed flame transport in a SI engine using simultaneous dual-plane OH-LIF and stereoscopic PIV

B. Peterson[1], E. Baum[2], A. Dreizler[2], B. Böhm[2]

[1]School of Engineering, Institute for Multiscale Thermofluids, University of Edinburgh, The King's Buildings, Mayfield Road, Edinburgh, EH9 3BF, UK

[2]Fachgebiet Reaktive Strömungen und Messtechnik (RSM), Technische Universität Darmstadt, Otto-Berndt-Straße 3, 64287 Darmstadt, Germany

**Corresponding author:** Brian Peterson, The King's Buildings, Mayfield Road, Edinburgh, EH9 3BF, Scotland, UK, Fax: +44 (0)131 650 6554, E-mail: brian.peterson@ed.ac.uk

**Abstract**

Understanding the detailed flame transport in IC engines is important to accurately predict ignition and combustion phasing, rate of heat release and assess engine performance. This is particularly important for RANS and LES engine simulations, which often struggle to accurately predict flame propagation and heat release without first adjusting model parameters. Detailed measurements of flame transport in technical systems are required to guide model development and validation.

This work introduces an experimental dataset designed to study the detailed flame transport and flame/flow dynamics for spark-ignition engines. Simultaneous dual-plane OH-LIF and stereoscopic PIV is used to acquire 3D measurements of unburnt gas velocity ($\vec{U}_{gas}$), flame displacement speed ($S_D$) and overall flame front velocity ($\vec{U}_{Flame}$) during the early flame development. Experiments are performed in an optical engine operating at 800 and 1500 RPM with premixed, stoichiometric isooctane-air mixtures. Analysis reveals several distinctive flame/flow configurations that yield a positive or negative flame displacement for which the flame progresses towards the reactants or products, respectively. For the operating conditions utilized, $S_D$ exhibits and inverse relationship with flame curvature and a strong correlation between negative $S_D$ and convex flame contours is observed. Trends are consistent with thermo-diffusive flames, but have not been quantified in context of IC engines. Flame wrinkling is more severe at the higher RPM, which results in a broader $S_D$ distribution towards higher positive and negative velocities. Spatially-resolved distributions of $\vec{U}_{gas}$ and $S_D$ are presented to describe in-cylinder locations where either convection or thermal diffusion is the dominating mechanism contributing to flame transport. Findings are discussed in relation to common engine flow features, including flame transport near solid surfaces. Findings provide a first insight into the detailed flame transport within a technically relevant environment and are designed to support the development and validation of engine simulations.

## 1. Introduction

In many combustion systems, such as spark-ignition (SI) engines, combustion efficiency is largely determined by the speed at which a flame transverses through a fuel-air mixture. Accordingly, one of the overarching goals in combustion science is to understand the detailed mechanisms of flame transport. In this context, we are concerned with the speed at which a wrinkled flame front traverses a certain distance. In turbulent combustion, the velocity of the flame front is denoted as the vector sum between local convection ($\vec{U}_{gas}$) and thermal diffusion ($S_D$) in the flame normal direction ($\vec{n}$) [1]:

$$\vec{U}_{Flame} = \vec{U}_{gas} + \vec{n} \cdot S_D \quad (1)$$

$S_D$ is commonly referred to as the flame displacement speed and is a central quantity in the understanding and modelling of turbulent premixed combustion [2]. Numerous numerical and experimental studies have been performed to understand the intrinsic behavior of $S_D$ with respect to flame structure [e.g. 3-7], flame stretch (including strain and curvature) [e.g. 6-9], turbulence intensity [e.g. 10, 11], and pressure [e.g. 12]. These fundamental investigations are often performed for a well-defined flame geometry (e.g. Bunsen, spherical or flat flame) within a well-characterized flow environment. Findings from such laboratory-scale flames have revealed the intrinsic behavior of $S_D$ with respect to various physicochemical and turbulent flow properties, which has greatly expanded our knowledge of flamelet theory [13-14].

While detailed investigations in laboratory-scale flames have provided a comprehensive understanding of flame dynamics in well-defined flame/flow geometries, flame behavior in technical systems (e.g. SI engines) is far less understood. In SI engines, we are not only interested in the intrinsic behavior of $S_D$ at high pressures, but also the capacity of $\vec{U}_{gas}$ to provide a fast, efficient flame transport. Additionally, it is important to resolve the stochastic nature of $\vec{U}_{gas}$ because previous studies have demonstrated that local flow variations can lead to favorable or unfavorable flame development, including misfires [15-18].

A comprehensive understanding of flame transport in SI engines requires detailed experimental measurements resolving the coupling between chemical reaction and the turbulent flow. Laser diagnostics such as flame imaging coupled with velocimetry measurements are well-suited to achieve this goal. Several investigations have utilized simultaneous flame-flow imaging methodologies in internal combustion (IC) engines [15-20], but only a few investigations have studied flame transport in detail [21, 22]. Mounaïm-Rousselle et al. utilized high-speed laser tomography and particle image velocimetry (PIV) to resolve local 2D values of $S_D$ in a boosted optical SI engine [21]. Using a short laser pulse separation, Mie scattering images resolved a planar view of flame displacement, while 2D $\vec{U}_{gas}$ velocities were directly measured from PIV. Measurements evaluated spatially-averaged $S_D$ values as the flame progressed in time for different dilution levels.

Recognizing limitations in 2D measurements, Peterson et al. utilized a multi-planar approach to resolve local 3D flame displacement speeds in an optical engine [22]. The approach, originally proposed by Trunk et al. [23], utilized dual-plane laser induced fluorescence (LIF) imaging of the hydroxyl radical (OH) to resolve flame displacement on parallel planes. A 3D flame surface was constructed between each plane, providing the flame normal direction in 3D space. Stereoscopic PIV (SPIV) was simultaneously applied with OH-LIF to measure the convection of the reconstructed flame surface. Measurements successfully resolved distributions of $\vec{U}_{gas}$ and $S_D$ during early flame propagation. Findings revealed a broad distribution of $S_D$ and the importance of measuring flame/flow quantities in 3D, which testified to the complex nature of flame development during the early stages of combustion.

The multi-dimensional, multi-parameter measurements presented in [22] provide a unique opportunity to investigate detailed flame transport and flame/flow dynamics in IC engines. This is particularly important for the development of predictive combustion models utilized in CFD engine simulations. While many sophisticated flame propagation models exist, many models struggle to accurately describe heat release without first adjusting model parameters. Overall, there is a lack of detailed measurements describing flame transport, which limits the development of more predictive flame models. In turn, this limits the use of CFD to be an effective design tool.

The work presented in this study expands on the measurements in [22] to introduce an experimental dataset designed to study detailed flame transport and flame/flow dynamics in IC engines. Measurements of $S_D$, $\vec{U}_{gas}$, and $\vec{U}_{Flame}$ are presented for premixed, stoichiometric isooctane-air mixtures at two different engine speeds of 800 and 1500 RPM. Measurements are acquired at a single crank-angle degree (°CA) during the early flame development phase when less than 5% of the mixture is consumed. Findings reveal distinctive flame/flow configurations that yield a positive or negative flame displacement for which the flame progresses towards the reactants or products, respectively. Locally resolved $S_D$ is evaluated with respect to 2D flame curvature along flame surfaces to describe the flame behaviour that is potentially responsible for yielding negative $S_D$ values. Finally, spatially resolved distributions of $S_D$ and $\vec{U}_{gas}$ are presented for each engine speed to describe in-cylinder locations where either thermal diffusion or convection is the more dominating mechanism contributing to flame transport.

## 2. Experimental

### 2.1 Engine

Experiments were performed in a single-cylinder optically accessible SI engine (AVL). The engine is equipped with a 4-valve pentroof cylinder head, centrally-mounted spark plug, and quartz-glass cylinder and flat piston. Details of the engine are described in [24, 25]. Measurements were performed at two engine speeds (800 and 1500 RPM) with homogeneous, stoichiometric isooctane-air mixtures from port-fuel injection. Port-fuel injection took place 1.4 m upstream of the engine and the isooctane fuel was pre-vaporized upon spark timing. Operating parameters, shown in Table 1, were chosen to mimic low-load engine operation. This operation is technically relevant as it can be prone to combustion instabilities [15-17]. Engine speed and air intake conditions were chosen to agree with the comprehensive velocimetry database for the *non-reacting* flow that is associated with this engine [24-30]. The intake temperature, however, is slightly higher in this study than in the aforementioned database because the port-fuelled section of the intake pipe was heated to promote sufficient fuel evaporation.

*Table 1: Engine operating conditions. °CA are referenced to top-dead-center compression.*

| | |
|---|---|
| Engine speed | 800, 1500 RPM |
| Intake Press., Temp. | 0.95 bar, 323 K |
| Fuel, lambda | $C_8H_{18}$, 1.0 |
| Ignition timing | 800 RPM: −19° bTDC<br>1500 RPM: −27° bTDC |
| IMEP, COV | 800 RPM: 5.2 bar, 1.7%<br>1500 RPM: 5.8 bar, 1.8% |
| OH-LIF, SPIV Image Timing | 800 RPM: −14° CA<br>1500 RPM: −18° CA |

Engine operating conditions provided repeatable thermodynamic conditions at ignition and image timings. Figure 1 shows the in-cylinder pressure trace for each engine speed. Before ignition, the in-cylinder pressure is lower for 800 RPM because there is more time for heat loss, which subsequently reduces the thermodynamic state. As a result, peak pressures are higher for 1500 RPM than 800 RPM.

Ignition timing was chosen such that the OH-LIF and SPIV measurement quality was optimized. As the piston approached top-dead-center (TDC), the field-of-view (FOV) became much smaller, particularly for SPIV because the cameras operated at an angle relative to the imaging plane.

Additionally, laser reflections at surfaces reduced SPIV quality near TDC, while higher gas temperatures near TDC reduced OH-LIF signal-to-noise ratios. At 800 RPM, measurements were optimized at −14 °CA (crank-angle degrees; referenced to TDC compression). Ignition at −19°CA provided sufficient time (5°CA or 1040μs) to initiate a flame kernel such that images captured the early flame kernel development when less than 5% of the mixture was consumed and the in-cylinder pressure was insensitive to the flame development. At 1500 RPM, an imaging timing of −18°CA provided a similar in-cylinder pressure to −14°CA at 800 RPM (11.1 bar). Ignition at −27°CA provided 9°CA (888μs) for flame development, which provided similar flame imaging conditions to those at 800 RPM (i.e. flame consumed less than 5% of the mixture and in-cylinder pressure was insensitive to flame development at image timing). These ignition timings provided optimal combustion stability (COV of IMEP < 2%) with indicated mean effective pressure (IMEP) greater than 5 bar.

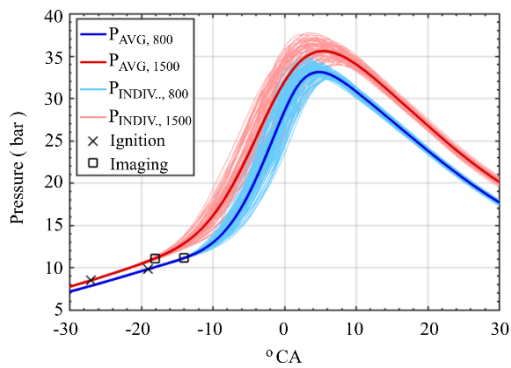

*Figure 1: In-cylinder pressure traces at 800 and 1500 RPM. Individual pressure traces encompass 100 cycles at each RPM.*

*2.2 Diagnostics*

The experimental setup, shown in Fig. 2, is the same as that used in [22]. A frequency-doubled Nd:YAG dual-cavity laser (Edgewave, INNOSLAB, 532 nm) operating at 4.8 kHz was used for SPIV measurements. Laser light passed through a set of focusing optics to form a laser sheet of 0.5 mm thickness. The sheet reflected off a 45° mirror in the crankcase and passed through the quartz glass piston to provide a sheet centered vertically with the cylinder axis and bisected the spark plug center electrode. Two CMOS cameras (Phantom V.711, double-frame exposure), placed on each side of the engine in Scheimpflug arrangement, imaged Mie scattering off chemically inert boron nitride (BN) particles (3 μm diameter), which were seeded into the intake flow. The cameras were mounted with an off-normal angle ($\gamma = 14°$) to the LIF cameras and imaged onto a region of 25 x $H$ mm$^2$, where $H$ was determined by the piston height.

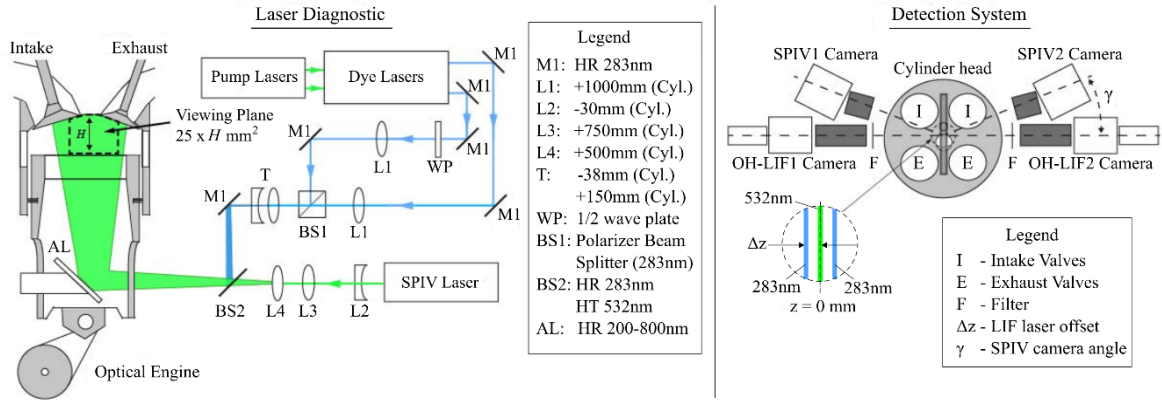

*Figure 2: Experimental setup of the multi-plane detection system in the optical engine.*

The hydroxyl radical (OH) was imaged simultaneously in two parallel, vertical planes. Two dye laser systems (Sirah, Precision Scan) operating with Rhodamine 6G were pumped by two separate double-pulsed Nd:YAG laser systems (Spectra Physics, PIV400, 532 nm). Dye lasers were tuned to 282.9 nm to excite the $Q_1(6)$ line of the A-X(1-0) transition of OH. Each laser system provided two UV laser pulses (24 mJ/pulse) that were temporally separated by $\Delta t_{LIF}$.

A half-wave plate and polarizing beam splitter were used to combine UV beams into two parallel paths. The beams travelled through focusing optics, providing two parallel UV light sheets (0.2 mm thickness) offset by 1 mm. A dichroic beam splitter combined the UV sheets with the path of the SPIV laser and directed the UV light into the engine. UV light sheets were offset by $\Delta z = \pm 0.5$ mm on each side of the PIV light sheet (see Fig. 2). Sheet separation and thickness were monitored by a UV-sensitive beam monitor (DataRay). Fluorescence emission from each laser sheet passed through high-transmission band-pass filters (UV-B) and was imaged onto two separate image intensifiers (LaVision, High-speed IRO) coupled to 14-bit CCD cameras (LaVision, ImagerProX, double-framed exposure) arranged on each side of the engine. Images were focused onto the same 25 x $H$ mm² FOV as the SPIV measurements. Spontaneous OH* chemiluminescence and flame luminosity were suppressed by gating each intensifier to 300 ns. The projected *pixel* resolution of both LIF detection systems was 20 μm, while the *spatial* resolution, determined by a Siemens-stern (contrast transfer function) was 80 μm.

An optical crank-angle encoder (AVL) was used to synchronize all lasers and cameras to the engine. The LIF systems were synchronized by a programmable timing unit (LaVision, PTU) to provide images at a fixed °CA. Each double-pulsed LIF system operated at 10 Hz providing two temporally resolved LIF images ($t_0, t_0 + \Delta t_{LIF}$) in each plane. The UV pulse separation ($\Delta t_{LIF}$) was set to 50 μs and 25 μs for 800 and 1500 RPM, respectively. For each RPM, $\Delta t_{LIF}$ was optimized to detect clear movement of the flame front. LIF images were recorded at a fixed °CA after ignition; at 800 RPM, LIF images were recorded at −14°CA, while at 1500 RPM LIF images were recorded at −18°CA. The UV pulses for each laser system were offset by 400 ns to avoid cross talk between LIF images in each plane.

SPIV lasers and cameras operated at 4.8 kHz to measure the three-component (3C) velocity field. At 800 RPM, SPIV measurements were recorded every °CA from −20° to −2°CA, while at 1500 RPM measurements were recorded every *two* °CAs from −28° to −4°CA. The pulse separation for SPIV ($\Delta t_{PIV}$) was 25 μs and 10 μs for 800 and 1500 RPM, respectively. The selection of $\Delta t_{PIV}$ was chosen to achieve a maximum particle shift of ¼ the final interrogation window size. The values of $\Delta t_{PIV}$ and $\Delta t_{LIF}$ differ because each technique detects the movement of separate substances (BN particle vs. OH layer) and because of the fundamental differences between SPIV and LIF processing approaches. When combined with LIF imaging, the first SPIV laser pulse was triggered $\Delta t_{PIV}/2$ after the first UV laser pulse. LIF images were recorded every other cycle for a 200-cycle sequence, while SPIV images were recorded for 200 consecutive cycles.

*2.3 Data processing and flame speed calculation*

The absolute velocity of the flame front ($\vec{U}_{Flame}$) is defined as the sum of the local unburnt convection velocity and the flame displacement speed relative to the flow in the flame-normal direction. This is shown schematically in Eq. (1) and in Fig. 3.

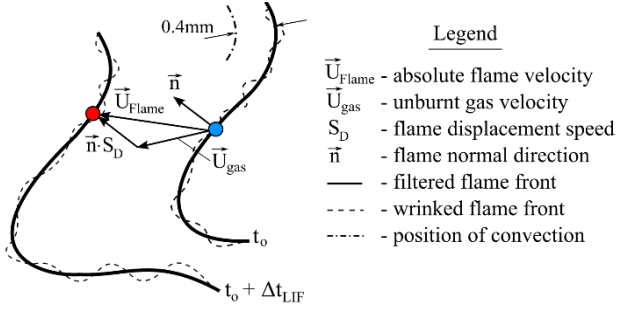

Figure 3: Vectorial schematic of local flame transport.

The image process procedure utilized to reconstruct a 3D flame surface and determine $S_D$ is described in [22] and is summarized below. A non-linear diffusion filter [31] based on an anisotropic operator splitting in combination with a Canny edge detection algorithm was used to detect the maximum LIF gradient between the burnt and unburnt gas. This maximum gradient was identified as the flame front. It is recognized that OH is a post-flame gas. Thus, the maximum OH gradient is located behind the physical flame front reaction-zone. Calculations using a 1D flamelet simulation [32] suggested that the laminar flame thickness is $\delta_L$ = 45 μm at the engine operating conditions (11.1 bar, ~550K). This thickness is in between the *pixel* resolution (20 μm) and the *spatial* resolution (80 μm) of the LIF detection systems. Within the short time duration of $\Delta t_{LIF}$, we do not expect large deviations between the progression of the reaction-zone and the progression of the maximum OH gradient. It is therefore anticipated that the maximum OH gradient is suitable to determine $\vec{U}_{Flame}$ and $S_D$ in this work.

A NURBS spline interpolation [33] was used to construct a 3D flame surface between the flame contours identified in each OH plane. A patch diffusion algorithm [34] removed numerical noise from the constructed surface. The 3D flame surface is projected through the SPIV plane providing the local flame-normal angle at z = 0 mm. The convection velocity was extracted 0.4 mm in front of the flame surface (flame normal direction) on the SPIV plane at $t_0$. This location is represented by the dotted line indicated as 'position of convection' in Fig. 3. Calculations using a 1D laminar flamelet simulation [32] demonstrated that the 0.4 mm distance was sufficient to avoid thermophoresis effects.

The 3D flame displacement speed was determined by transporting the position of the flame front at $t_0$, z = 0 mm by the local 3C velocity. The remaining distance between the transported contour and the nearest point on the reconstructed flame surface at time $t_0 + \Delta t_{LIF}$ in the flame normal ($\vec{n}$) is representative of the local displacement speed, $S_D$. Local values were determined for individual points spaced 20 μm along the flame contour (i.e. *pixel* resolution) on the SPIV plane. The limited *spatial* resolution (80 μm LIF) is greater than the laminar flame thickness ($\delta_L$ = 45 μm). The detected flame front in the measurements is therefore recognized as a spatially filtered quantity.

SPIV images were processed with a commercial software (LaVision, DaVis 8.1). Spatial calibration and dewarping of the SPIV images were accomplished with a 3D target (LaVision, Type7). Self-calibration was accomplished from 200 Mie scattering images before ignition and provided a remaining average pixel disparity less than 0.01 pixels. Image cross-correlation and vector calculations were performed with a decreasing window size multi-pass algorithm. The final interrogation window size was 24 x 24 with 75% overlap, providing a 0.15 mm vector spacing.

Due to the limited spatial and temporal resolutions of each diagnostic, the quantities of $S_D$, $\vec{U}_{Flame}$, and $\vec{U}_{gas}$ reported in this work are considered to be *filtered* quantities. The effect of the limited measurement resolutions are discussed within *Sect. 2.4*.

*2.4 Measurement uncertainty*

A detailed uncertainty and sensitivity analysis has been presented in [22] and a brief summary is provided here. The most influential parameters affecting measurement uncertainty are the LIF resolution and the accuracy of the unburnt gas velocity. The LIF detection systems, having a *spatial* uncertainty of 80 μm, could result in a *maximum* possible offset (i.e. uncertainty) of $\delta_{max}$ = 160 μm between the flame surfaces identified at $t_0$ and $t_0 + \Delta t_{LIF}$. Assuming this offset detection is Gaussian distributed, an offset of 1σ between flame surfaces is $\delta_{1\sigma}$ = 53 μm. The resulting uncertainty of $\vec{U}_{Flame}$ and $S_D$ associated with the LIF detection limits is $\psi_{1\sigma,max} = \delta_{1\sigma,max}/\Delta t_{LIF}$. At 800 RPM, $\psi_{1\sigma}$ = 1.1 m/s and $\psi_{max}$ = 3.2 m/s, while at 1500 RPM $\psi_{1\sigma}$ = 2.1 m/s and $\psi_{max}$ = 6.4 m/s.

Uncertainty of SPIV measurements are dependent on several parameters. Analysis of the SPIV particle response time (see Appendix A) demonstrates that BN particles with diameter of 3 μm will accurately follow the engine flow in this study. With an optimized experimental setup (i.e. optimized camera angles, depth of field, seeding density) and sophisticated processing algorithms (e.g. cross-correlation, adaptive PIV with variable interrogation window size and shape [35], high accuracy mode for final passes), SPIV measurements have an estimated uncertainty of ≤ 10%. This would correspond to a *maximum* $\vec{U}_{gas}$ uncertainty of 1.2 m/s (800 RPM) and 2.1 m/s (1500 RPM).

Given the aforementioned uncertainties, a root mean square estimation for $S_D$ provides a 1σ uncertainty of ±1.5 m/s and ±3.0 m/s at 800 and 1500 RPM, respectively. *Maximum* uncertainties yield ±3.4 m/s and ±6.7 m/s at the respective RPMs.

Peterson et al. also performed a rigorous sensitivity analysis on the calculation of $S_D$ [22]. This sensitivity analysis included variations of SPIV spatial resolution, location of $\vec{U}_{gas}$ relative to the flame surface, and laser sheet spacing (Δz). Variation in SPIV resolution and $\vec{U}_{gas}$ location yielded an average deviation of $\Delta S_D$ = 0.8 m/s, which did not alter the $S_D$ distribution. Experiments performed with Δz = 0 mm (i.e. single-plane measurements) revealed that differences in each LIF detection system can yield a maximum artificial normal flame angle of β ± 12°. This would yield a maximum bias of $\Delta S_D$ of 0.4 m/s. Additional experiments with Δz = 0.25 and 0.75 mm revealed similar $S_D$ distributions as those from Δz = 0.5 mm. Assuming that out-of-plane curvature is similar to in-plane curvature, it was argued that the linear reconstruction method with Δz = 0.5 mm is suitable to capture the local 3D flame structure. In addition to a rigorous conditional sampling analysis, it was concluded that $S_D$ distributions using this method are credible within the reported uncertainty.

### 3. Statistical distributions

*3.1 Flow field and burnt gas*

Statistical distributions of the flow field and burnt gas locations are presented to provide an overview of the turbulent flow and flame environment in the SI engine. Figure 4 shows the ensemble-averaged flow field and the probability distribution of the burned gas at each RPM. Flow fields are shown before ignition timing and at the °CA for which OH-LIF measurements were acquired. Velocity statistics are based on 200 consecutive engine cycles. Streamlines are used to show the flow direction, while the color-scale depicts the 3C velocity magnitude. Burnt gas probability maps are shown at OH-LIF timing and were constructed from binarized LIF images identifying the burnt gas. These distributions are based on 100 engine cycles. Ensemble-averaged velocity vectors are overlaid onto the burnt gas distributions.

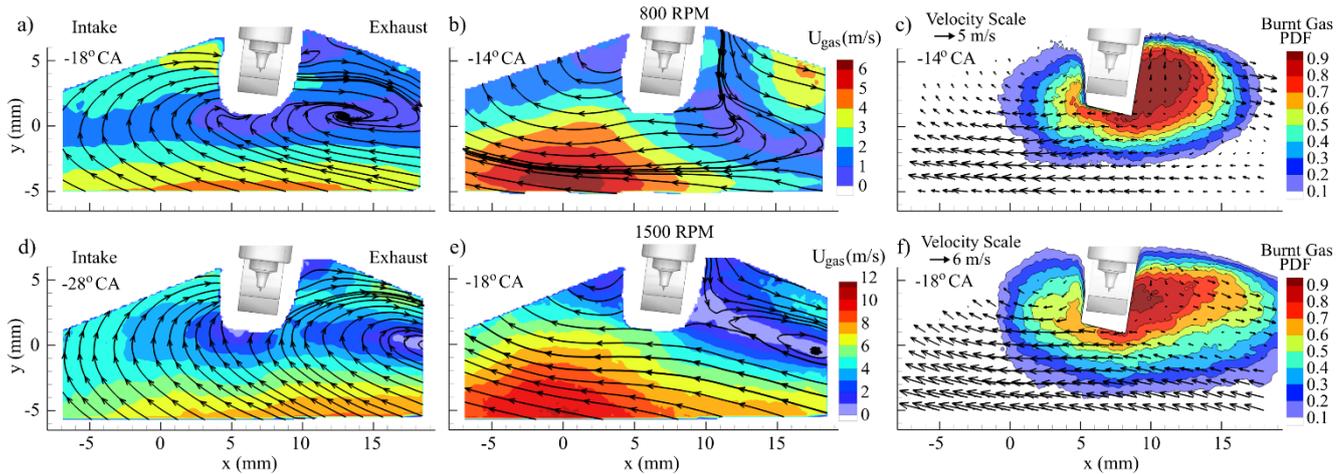

*Figure 4: Ensemble-average flow-fields and burnt gas PDFs at 800 RPM (top) and 1500 RPM (bottom). Left: flow field before spark timing. Middle: flow field at OH-LIF image timing. Right: burnt gas PDFs and ensemble-average flow field (every 6$^{th}$ vector shown). Velocity statistics are based on 200 PIV images. PDF statistics are based on 100 LIF images.*

Directly before spark timing, a clockwise tumble motion exists for both RPMs. At 1500 RPM, however, velocity magnitude is greater and the tumble center is shifted approximately 6 mm further to the right within the FOV. Differences of the tumble center location with RPM have been reported by the authors in previous work [28]. The cause of this shift is not investigated in this study.

At the OH-LIF image timing, the flow patterns for each RPM are qualitatively similar, while higher velocity magnitudes are shown for 1500 RPM. The flow patterns exhibit a "*sweeping*" like flow motion from right-to-left in the lower-half of the FOV. This *sweeping* flow motion is primarily due to the piston movement towards TDC. For both RPMs, the highest velocities in the FOV exist southwest of the spark plug and are symmetrically centered near the cylinder axis (i.e. x = 0 mm). The upper-half of the FOV exhibits lower velocity magnitude, particularly to the right of the spark plug where the tumble center is located. At 1500 RPM, the tumble center location is captured within the FOV, while at 800 RPM, SPIV images preceding −14°CA suggest that the tumble center is located directly next to the spark plug and is not as clearly identified in the FOV.

The burnt gas distributions exhibit several similarities between 800 and 1500 RPM. In brief, a flame develops radially outward from the spark plug with a greater tendency to propagate towards the exhaust side owing to the predominant left-to-right flow direction near the spark plug prior to ignition. At 1500 RPM, however, the enflamed gas region is larger and extends further towards the exhaust side of the cylinder. Before ignition, the tumble center location being further from the spark plug and the overall higher velocity magnitudes are such that the flow at the spark plug is more strongly directed towards the exhaust side of the cylinder for 1500 RPM than for 800 RPM. Higher velocity magnitudes at 1500 RPM transport the flame further from the spark plug, which is also responsible for the larger enflamed area in the negative y-region.

For completion, RMS velocity fields are shown in Fig. 5 to characterize the turbulence associated with the in-cylinder flow. RMS fields are constructed from Reynolds decomposition (200 cycle statistic) and considers all three velocity components. Selected iso-contours of the enflamed gas probability distribution are overlaid onto the RMS field. RMS velocities in the burnt gas region ranges from 2.5 – 4.0 m/s for 800 RPM and 5.0 – 8.0 m/s for 1500 RPM. Putting this into perspective with the laminar burning velocity ($S_L$= 0.36 m/s) gives $(u'/S_L)_{800\ RPM}$ = 6.9 – 11.1 and $(u'/S_L)_{1500\ RPM}$ = 13.8 – 22.2. The Reynolds decomposition method accounts for flow turbulence and flow variations from cycle-to-cycle. Although the reported $u'/S_L$ values are likely overestimated, they provide a relative comparison of the turbulent flow at each RPM. The reader is referred to our earlier work for further discussion on turbulence characterized by Reynolds decomposition and its relation to *instantaneous turbulence* using tomographic PIV [28].

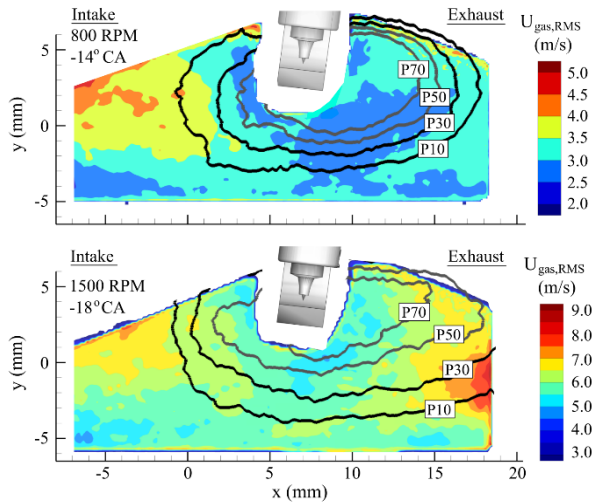

Figure 5: RMS velocity distribution at 800 and 1500 RPM. RMS field shown at OH-LIF image timing and based on 200 PIV images. Selected PDF burnt gas contours are overlaid onto the RMS field.

Overall, the velocity and enflamed gas distributions presented in Figs. 4 and 5 demonstrate a similar initial flame development between 800 and 1500 RPM at the selected image timings. Thus, with similar thermodynamic conditions and only minor differences in flow patterns between 800 and 1500 RPM, the selected operating parameters provide a practical environment to study flame propagation for different convective velocity magnitudes and turbulence levels.

## 3.2 PDF distributions: $S_D$, $\vec{U}_{gas}$, and $\vec{U}_{Flame}$

In this work, values of $S_D$, $\vec{U}_{gas}$, and $\vec{U}_{Flame}$ are spatially measured along flame surfaces. PDFs of these velocities are presented in Fig. 6 to describe the range of velocities resolved for each engine operation. Ensemble-average and standard deviation are reported within each subplot. The data is composed from 100 engine cycles, which encompasses 78,030(79,223) data points at 800(1500) RPM.

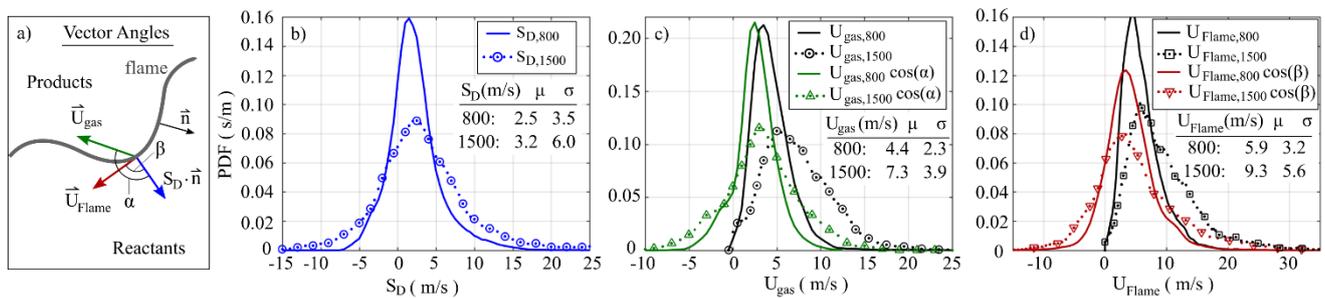

Figure 6: (a) schematic of vector angles with respect to flame normal. (b-d): PDF of $S_D$, $\vec{U}_{gas}$, and $\vec{U}_{Flame}$. $\vec{U}_{gas} \cos(\alpha)$ and $\vec{U}_{Flame} \cos(\beta)$ distributions are also reported to reveal velocity distributions relative to the flame normal. Statistics are based on 100 engine cycles, yielding 78,030 and 79,223 data points at 800 and 1500 RPM, respectively.

For each velocity, the *magnitude* and the *direction* relative to the flame normal are of interest. $S_D$ is always normal to the flame surface. Thus, the sign of $S_D$ will automatically determine its *magnitude* and *direction* relative to the flame surface; a positive $S_D$ indicates transport toward the reactants, while a negative $S_D$ indicates transport towards the products. The latter is identified as a *negative*

*flame speed*. This phenomenon is discussed further in *Sects. 4.1* and *4.2*. $\vec{U}_{gas}$ and $\vec{U}_{Flame}$ do not have a fixed angle to $\vec{n}$. Within this manuscript, $\vec{U}_{gas}$ and $\vec{U}_{Flame}$ values are referred to as *velocity magnitude* and unless otherwise stated, they will always exhibit a *positive* value. $\vec{U}_{gas}\cos(\alpha)$ and $\vec{U}_{Flame}\cos(\beta)$ velocities are used to indicate the *direction* relative to the flame normal. These velocities can be either positive or negative depending on the vector angle; positive velocities indicate transport towards the reactants, while negative velocities indicate transport towards the products. The vector angles $\alpha$ and $\beta$ are calculated from the 3D projection of $\vec{U}_{gas}$ and $\vec{U}_{Flame}$ onto $\vec{n}$. Distributions of $\vec{U}_{gas}\cos(\alpha)$ and $\vec{U}_{Flame}\cos(\beta)$ are also included in Fig. 6.

Referring to Fig. 6, as engine speed increases from 800 to 1500 RPM, the velocity distributions broaden. $S_D$, $\vec{U}_{gas}\cos(\alpha)$, and $\vec{U}_{Flame}\cos(\beta)$ distributions all exhibits longer tails towards both higher positive and negative velocities, while $\vec{U}_{gas}$ and $\vec{U}_{Flame}$ distribution show a pronounced shift towards higher velocity magnitudes as engine speed increases. At 800 RPM, 11.2% of the data reports negative $\vec{U}_{Flame}\cos(\beta)$ values, while 19.9% and 13.7% of the data report negative $S_D$ and $\vec{U}_{gas}\cos(\alpha)$ values, respectively. At 1500 RPM, these negative distributions increase to 19.5%, 25.5% and 21.9% for $\vec{U}_{Flame}\cos(\beta)$, $S_D$ and $\vec{U}_{gas}\cos(\alpha)$, respectively. Thus, while overall flame progression is faster at 1500 RPM (i.e. higher positive velocities), there also appears to be a tendency for stronger local flame recession (i.e. negative velocities) at the higher engine speed.

Although the PDFs presented in Fig. 6 provide a general overview of the flame transport velocities, they do not reveal the physical mechanisms attributing to fast/slow flame progression. With such broad distributions of flame transport, it is important to understand the local combination of $S_D$ and $\vec{U}_{gas}$ velocities that appropriately yield favourable or unfavourable flame progression. The remainder of the paper presents analysis that investigates the local flame transport mechanisms that describe the PDF distributions presented in Fig. 6.

## 4 Results

This section presents local distributions of $S_D$, $\vec{U}_{gas}$, and $\vec{U}_{Flame}$ resolved along flame contours to describe local flame transport. *Section 4.1* discusses unique flame/flow interactions that yield positive or negative flame displacement for which the flame progresses towards the reactants or products, respectively. *Section 4.2* presents locally resolved $S_D$ velocities with respect to 2D flame curvature to discuss flame dynamic behaviour that increases or decreases $S_D$ velocities. Finally, findings presented in *Sect. 4.3* describe the in-cylinder locations where $S_D$ or $\vec{U}_{gas}$ is the dominating mechanism responsible for flame transport.

Within the following discussion we use the words 'convection' and 'thermal diffusion' to refer to $\vec{U}_{gas}$ and $S_D$, respectively. It is acknowledged that physical mechanisms such as mass diffusion, flow dilatation, and surface density can also contribute to $S_D$. The Lewis number (*Le*) for the $C_8H_{18}$-air mixtures in this work is 1.99 (see Sect. 4.2). It is anticipated that thermal diffusion will have a more dominant role than mass diffusion. Several investigations (e.g. [36-38]) have shown flow dilatation can influence flame transport behavior, particularly for statistically flat flames. Chakraborty et al. [37] demonstrated that dilatation effects are less significant for spherical flame kernels than flat flames. Flame kernel development in the engine is more consistent to the spherical flame growth than a flat flame. However, there are many aspects in engine environments that previous studies have not addressed; e.g. dilatation (and surface density) effects at high pressure, high temperature, and various deviations from perfectly spherical flame propagation. While several arguments suggest that $S_D$ measured in our experiments is primarily depicted by thermal diffusion, further evidence is required to understand the role of mass diffusion, flow dilatation, and surface density on $S_D$ in engine environments. In this work, however, we believe that 'thermal diffusion' is a reasonable (and simple) terminology to describe $S_D$ and will be used hereafter.

*4.1 Flame/flow interactions*

This section evaluates flame/flow interactions that describe the local flame displacement. This analysis is first presented for an individual engine cycle at 1500 RPM. Figure 7 shows the OH-LIF images on each imaging plane for this individual engine cycle. The flame front, identified by the non-linear diffusion filter and Canny edge detection, is superimposed onto the LIF images. The flame front is well resolved along the OH-LIF image with the exception of the left-hand side for the z = –0.5mm plane. At time $t_o$ a weak OH-LIF signal exists outside the detected enflamed boundary, while at time $t_o + \Delta t_{LIF}$ the OH-LIF signal is stronger and is considered by the filter and detection algorithms. Images suggest that a flame from the –z direction enters the z = –0.5mm plane, causing the differences in the detected flame front. Such effects would lead to systematic errors in $\vec{U}_{Flame}$ and $S_D$ (e.g. $S_D$ > 50 m/s). This is a limitation of the planar techniques utilized. Such regions of large discrepancy were easily identified via visual inspection during the filtering and detection steps. Beyond visual inspection, the calculation of $S_D$ was suspended upon calculation of $S_D$ > 40 m/s (i.e. values greatly exceeding maximum values shown in Fig. 6). This further identified regions with large contour discrepancy between each frame. Such regions (for all cycles) were removed and not considered in the flame transport analysis. For the cycle in Fig. 7, approx. 28% of the flame contour was removed. At 1500 RPM, 22 out of 100 cycles required a portion of the flame contour to be removed. The percentage of flame contour removed ranged from 6% to 31% and the average length removed was 16%. At 800 RPM, only 18 out of 100 cycles required a portion of the flame contour to be removed; the percentage of flame contour removed ranged from 6% to 26% and the average length removed was 12%.

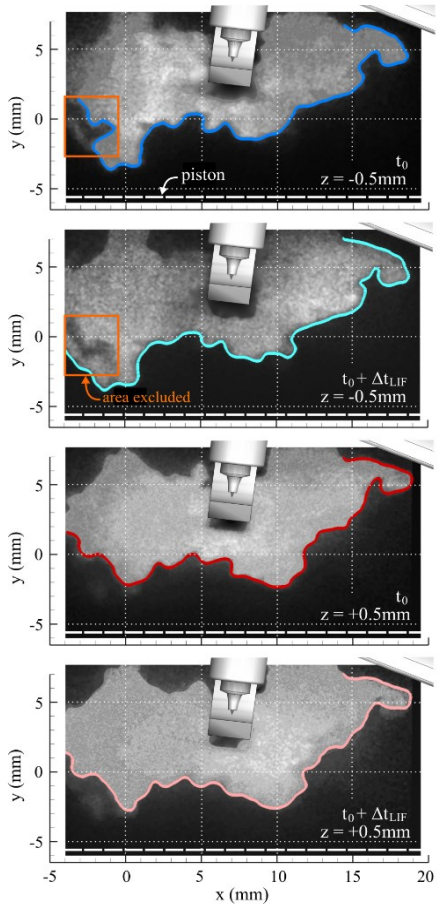

*Figure 7: OH-LIF images in respective planes and image timings for an individual engine cycle at 1500 RPM. LIF images shown have been corrected for laser profile and absorption. The identified reaction zones are indicated by colored lines. $\vec{U}_{Flame}$ and $S_D$ quantities are not calculated in highlighted region on the left.*

Figure 8a shows the reconstructed flame front (z = 0mm, $t_o$) overlaid onto the 2D3C flow field for the individual cycle. The flow field (every 6$^{th}$ vector shown) exhibits many similarities to the ensemble-average flow field in Fig. 4. The flow resembles the *sweeping* flow motion from right-to-left. The highest velocity magnitudes (≥ 10 m/s) are located in the left half of the FOV, while the region to the right of the spark plug exhibits lower velocities (≤ 6 m/s). The spark plug and pentroof cylinder head obstructed the view of the SPIV cameras, which were mounted at an angle ($\gamma$ = 14°) with the imaging plane. This restricted velocity measurements near the spark plug and cylinder head, while OH-LIF images were able to reveal the flame in these areas. This explains why the flame contour extends beyond the velocity field in the upper right corner of Fig. 8. Such regions (for all cycles) are not considered in the flame transport analysis.

For clarity and brevity, the detailed flame transport is described within a small flame segment highlighted by the rectangle shown in Fig. 8a. The flame transport resolved within this flame segment admirably describes different regimes of the PDF distributions shown in Fig. 6. The 3D flame surfaces temporally resolved at $t_o$ and $t_o+\Delta t_{LIF}$ are shown in Fig. 8b. Spatially resolved $\vec{U}_{Flame}$, $S_D$ and $\vec{U}_{gas}$ velocities (every 10$^{th}$ vector shown) along the flame surface ($t_o$) are displayed in Figs. 8 c,d,e respectively.

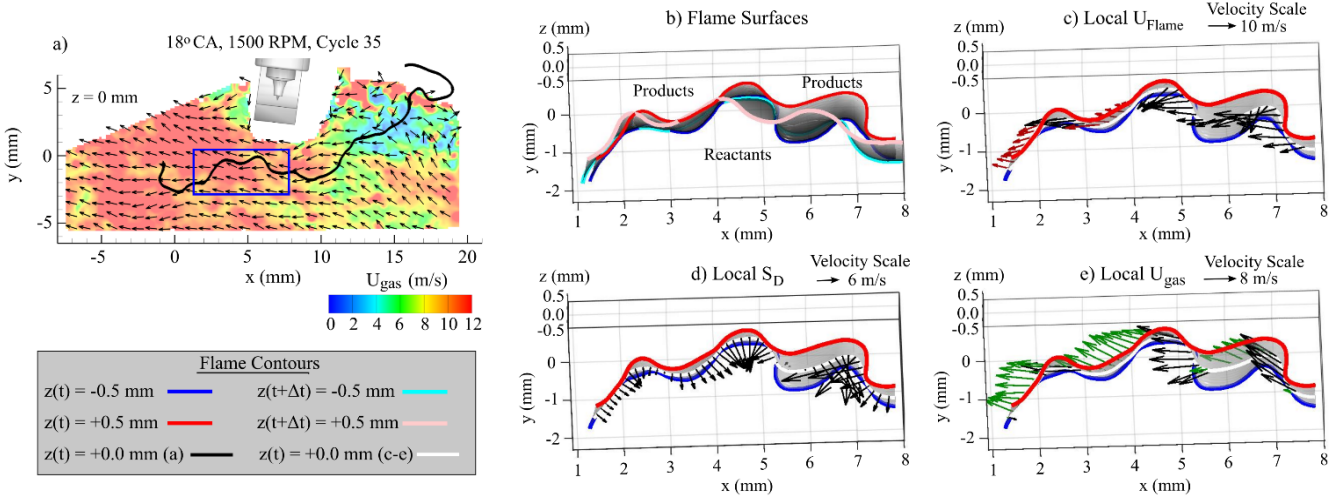

*Figure 8: (a) 2D flame contour (z = 0mm) overlaid onto the SPIV field for the individual cycle shown in Fig. 7. Blue rectangle highlights the flame segment shown in (b-e). (b) 3D flame surfaces at $t_o$ and $t_o+\Delta t_{LIF}$, (c) local $\vec{U}_{Flame}$ distribution; locations where the flame is transported to the products are highlighted by red vectors. (d) local $S_D$ distribution, (e) local $\vec{U}_{gas}$ velocities; locations where $\alpha > 90°$ are highlighted by green vectors.*

The flame surfaces shown in Fig. 8b reveal that the flame is not transported uniformly and, in some locations, the flame surface is transported in the direction of the products, rather than the reactants. This can be determined by evaluating $\vec{U}_{Flame}\cos(\beta)$ along the flame contour in Fig. 9. Figure 9 reports the velocity values and vector angles calculated along the flame segment. Flame contour points in Fig. 9 begin at the left-most position of the flame segment and walk along the flame segment in equally spaced points. When the vector angle angle ($\beta$) between $\vec{U}_{Flame}$ and $\vec{n}$ is greater than 90°, $\vec{U}_{Flame}\cos(\beta)$ is negative and the flame is displaced towards the products. Locations where the flame is transported towards the products are highlighted by red vectors in Fig. 8c and grey shaded regions in Fig. 9. Figure 9 reveals that $\beta$ can be as high as 150° such that the local flame propagation can be strongly directed towards the products.

$S_D$ velocities (Fig. 8d) are normal to the flame surface and span the range $–2.5 \leq S_D \leq 25$ m/s. A negative $S_D$ occurs when the convection velocity transports the flame contour further than the flame surface at time $t_o+\Delta t_{LIF}$. As a result, $S_D \cdot \vec{n}$ is in the direction of the products rather than reactants [22, 23]. Negative $S_D$ velocities are discussed further in *Sect. 4.2*. Within this flame segment however, $S_D$ is primarily positive, indicating that the flame propagates towards the reactants.

$\vec{U}_{gas}$ velocities shown in Fig. 8e, on the other hand, demonstrate that convection transports a portion of the flame towards the reactants and other portions of the flame towards the products. Unlike $S_D$, the vector angle between $\vec{U}_{gas}$ and $\vec{n}$ is not fixed and has a definitive role in the convective flame transport. Although the $\vec{U}_{gas}$ distribution for the flame segment in Fig. 8 exhibits a consistent flow direction, the flame surface exhibits severe wrinkling such that $\alpha$ significantly varies along the flame surface. $\vec{U}_{gas}\cos(\alpha)$ and $\alpha$ are evaluated along the flame segment and plotted in Fig. 9. Vector angles greater than 90° depict that the flame is transported towards the products, yielding a negative $\vec{U}_{gas}\cos(\alpha)$ value. In Fig. 8e, green vectors indicate locations where $\vec{U}_{gas}$ transports the flame contour towards the products, while black vectors reveal locations where $\vec{U}_{gas}$ transports the flame towards the reactants.

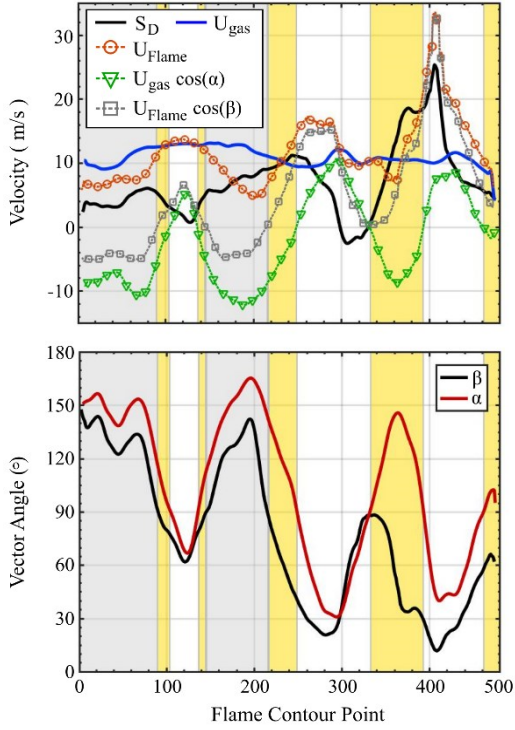

*Figure 9: Velocity and vector angle quantities along the flame contour in Fig. 8 b-e. Grey regions correspond to locations where $\vec{U}_{Flame}\cos(\beta) < 0$. Yellow regions correspond to locations where $\vec{U}_{gas}\cos(\alpha) < 0$, but $\vec{U}_{Flame}\cos(\beta) > 0$ such that the flame propagates towards the reactants.*

The flame/flow configurations, determined by $S_D$, $\vec{U}_{gas}$, and α, are responsible for the local $\vec{U}_{Flame}$ distribution. Locations where α < 90º, both $S_D$ and $\vec{U}_{gas}$ participate to transport the flame towards the reactants as long as $S_D > 0$. As $S_D$ and $\vec{U}_{gas}$ work in unison, this will yield higher (positive) $\vec{U}_{Flame}$ values than just $S_D$ alone. This can be seen in regions between x = 5.2 − 7.1mm (corresponding flame contour points 260−310 and 375-430, respectively, in Fig. 9) where α values are lowest (30º − 60º) and corresponding $\vec{U}_{Flame}\cos(\beta)$ values are highest (14.7 – 32.6 m/s). However, when α > 90º, the convective flow competes against the thermal diffusion such that $\vec{U}_{Flame}$ will be determined by larger magnitude between $\vec{U}_{gas}\cos(\alpha)$ and $S_D$. When α > 90º and $|\vec{U}_{gas}\cos(\alpha)| > S_D$, adverse convection will dominate thermal diffusion such that the flame will be transported towards the products, yielding $\vec{U}_{Flame}\cos(\beta) < 0$. However, if α > 90º, but $|\vec{U}_{gas}\cos(\alpha)| < S_D$, thermal diffusion will dominate the convective flow and the flame will progress towards the reactants as long as $S_D$ is positive. Such regions are highlighted in yellow in Fig. 9.

Flame/flow interactions for all cycles are described in Fig. 10 where the ratio of $\vec{U}_{gas}/S_D$ is analyzed with respect to the flame/flow angle, α. $\vec{U}_{gas}/S_D$ (y-axis) effectively evaluates the ratio of velocity *magnitudes* to determine which transport mechanism is greater. Recall that $\vec{U}_{gas}$ is considered to always be a positive value (i.e. purely velocity magnitude). $S_D$ is unique in that it is always normal to $\vec{n}$ and the sign of $S_D$ depicts the direction relative to $\vec{n}$; positive/negative $S_D$ indicates its transport is towards the reactants/products. Thus, $\vec{U}_{gas}/S_D < 0$ will only occur when $S_D < 0$. The x-axis ($\alpha$) evaluates the direction of $\vec{U}_{gas}$ relative to $\vec{n}$; $\vec{U}_{gas}$ transports the flame towards reactants when $\alpha < 90º$ and transports the flame towards the products when $\alpha > 90º$ (see Fig. 6a).

Data points in Fig. 10 are colored with respect to $\vec{U}_{Flame}\cos(\beta)$ values to identify flame/flow interactions that lead to a positive or negative overall flame displacement. In addition, data points corresponding to the highest (positive) 15% $\vec{U}_{Flame}\cos(\beta)$ values are highlighted to identify flame/flow interactions that lead to faster overall flame propagation. These data points correspond to $\vec{U}_{Flame}\cos(\beta)$ values exceeding 9.0 and 14.5 m/s at 800 and 1500 RPM, respectively.

The data points in Fig. 10 are distributed into quadrants I – IV, which are separated by x-y axes $\alpha = 90°$ and $\vec{U}_{gas}/S_D = 0$, respectively. Dashed lines located at $\vec{U}_{gas}/S_D = \pm 1$ separate regions where either convection or thermal diffusion are more dominant towards the overall flame displacement. Overall, several flame/flow relationships exist that lead to a positive or negative flame displacement. These relationships are shown by data points located in specific regions within each quadrant. Findings are qualitatively similar for 800 and 1500 RPM and are discussed below in terms of thermal diffusive ($S_D$) and convective ($\vec{U}_{gas}$) flame transport.

- Flame displacement is *positive* (i.e. $\vec{U}_{Flame}\cos(\beta) > 0$) when $\alpha < 90°$ and $S_D$ is a positive value. Such data is located in quadrant II. In this scenario, both convection and thermal diffusion work in unison to support flame displacement towards the reactants.
- Flame displacement is also *positive* beyond $\alpha > 90°$ when $0 < \vec{U}_{gas}/S_D < 1$. Such data is located in quadrant I. Although the convection velocity transports the flame towards the products in this scenario, the thermal diffusive velocity, directed towards that reactants, is always greater such that the overall flame displacement progresses towards the reactants.
- Flame displacement has the highest (positive) value when $\alpha < 90°$ and/or when thermal diffusion is more dominant than convection (i.e. $0 < \vec{U}_{gas}/S_D < 1$). This data is highlighted in dark blue in Fig. 10 and is primarily located in quadrant II. A small population of this data exists in quadrant I where $\alpha > 90°$. This, however, occurs when $\vec{U}_{gas}/S_D \leq 0.8$, indicating that although convection is directed towards the products, thermal diffusion directed towards the reactants is more dominant and exceeds the specified $\vec{U}_{Flame}\cos(\beta)$ threshold. As $\alpha$ decreases from 90° (quadrant II), data points exhibiting high $\vec{U}_{Flame}\cos(\beta)$ velocities also exist for $\vec{U}_{gas}/S_D$ values greater than unity. This occurs as the flame-normal convection velocity ($\vec{U}_{gas}\cos(\alpha)$) increases and contributes towards the positive flame displacement towards the reactants.
- Flame displacement is *negative* when $S_D < 0$ and $\alpha > 90°$. This data is located in quadrant IV. In this scenario, both convection and thermal diffusion transport the flame towards the products.
- Flame displacement is also *negative* when $S_D < 0$ and $\alpha < 90°$. Such data points are located in quadrant III where $0 > \vec{U}_{gas}/S_D > -1$. In this scenario, although convection transports the flame towards the reactants, thermal diffusion, which is directed towards the products, is the dominant velocity that leads to the negative flame transport.

In quadrants I and III, positive or negative flame displacement can exist in regions of $|\vec{U}_{gas}/S_D| > 1$ and is dependent on the values of $\vec{U}_{gas}/S_D$ and $\alpha$. In particular:

- In quadrant I, $\alpha > 90°$ such that flame displacement will be *positive* as long as $S_D > \vec{U}_{gas}\cos(\alpha)$. In this scenario, thermal diffusion transports the flame towards the reactants, while convection opposes $S_D$ and transports the flame towards the products. The flame will propagate towards the reactants as long as the flame-normal convection velocity is less than the thermal diffusion transport.
- As $\alpha$ increases in quadrant I, the convective transport, directed towards the products, will exceed the thermal diffusion transport directed towards the reactants. Flame displacement will become *negative* when $\vec{U}_{gas}\cos(\alpha) > S_D$.
- In quadrant III, $S_D$ transports the flame towards the products, while $\vec{U}_{gas}$ transports the flame towards the reactants. Flame displacement will be *positive* as long as the flame-normal convection velocity is greater than thermal diffusion, i.e. $\vec{U}_{gas}\cos(\alpha) > |S_D|$.

- As α increases in quadrant III, the convection velocity towards the reactants decreases. Flame displacement will become *negative* when thermal diffusion towards the products exceeds the flame-normal convection velocity, i.e. $|S_D| > \vec{U}_{gas}\cos(\alpha)$.

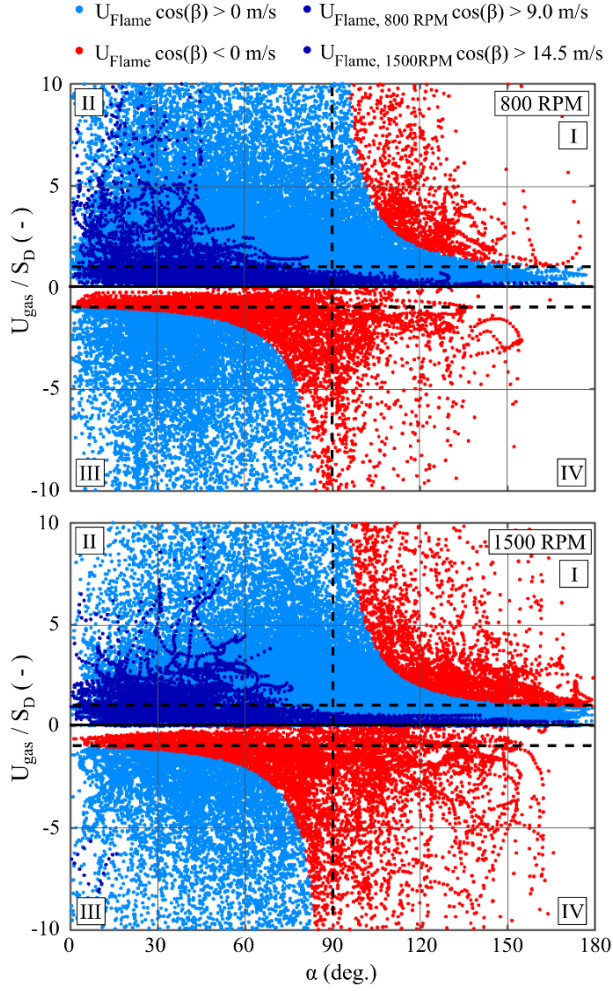

*Figure 10: Various flame/flow interactions for all cycles are evaluated within the quadrants of the $\vec{U}_{gas}/S_D$ vs. α diagram. Data points are colored with respect to positive and negative flame displacement. Flame/flow interactions leading to fast flame propagation are highlighted in dark blue. The percentage of data for positive and negative flame displacement within each quadrant is reported.*

| RPM | No. Pts | +I | +II | +III | -I | -III | -IV |
|---|---|---|---|---|---|---|---|
| 800 | 78,030 | 14.7% | 60.7% | 13.6% | 2.4% | 6.3% | 2.3% |
| 1500 | 79,223 | 23.8% | 42.9% | 13.8% | 6.3% | 7.5% | 5.7% |

Data Percentage

This analysis reveals the complex nature of flame transport and discusses the numerous flame/flow configurations that exist in the engine. The percentage of data for positive and negative $\vec{U}_{Flame}\cos(\beta)$ is reported in Fig. 10. As engine speed increases, the percentage of data in Q2 decreases, while the data percentage increases significantly for Q1 and Q4, revealing that a larger percentage of the flame/flow interactions occur with a larger flow/flame vector angle, α. This demonstrates the greater complexity of flame transport at higher RPMs and reveals one of the mechanisms leading to the higher occurrence of negative flame displacement at 1500 RPM.

4.2 Flame dynamics describing $S_D$ behavior

While the flame/flow interactions leading to positive or negative $\vec{U}_{Flame}\cos(\beta)$ values has been analyzed extensively, the large variation of $S_D$ values is less understood. In this section, $S_D$ is evaluated with respect to flame curvature ($\kappa$) to understand variations of flame speed along a flame contour in order to describe the $S_D$ distributions discussed in *Sects. 3.2* and *4.1*. Due to the limited z-resolution, local flame curvature is evaluated in 2D using the relation [4, 39]:

$$\kappa = \frac{\frac{dx}{ds}\frac{d^2y}{ds^2} - \frac{dy}{ds}\frac{d^2x}{ds^2}}{\left[\left(\frac{dx}{ds}\right)^2 + \left(\frac{dy}{ds}\right)^2\right]^{3/2}} \quad (2)$$

where s is the curvilinear coordinate. For this definition, curvature is considered positive (negative) when the flame is convex (concave) towards the reactants.

Figure 11 evaluates $S_D$ and $\kappa$ along sample sections of flame contours for two individual cycles that show different degrees of wrinkling. Both cycles are taken from the 1500 RPM dataset and the cycle shown in the bottom image is the same cycle discussed in Figs. 7-9. Figures 11 a,d show the OH-LIF images on the z = −0.5 mm plane and highlight the flame segments for which $S_D$ and $\kappa$ are analyzed in detail. Flame curvature is evaluated along the z = 0 mm flame contour (white contour, Figs. 11 b,e). Figures 11 c,f reveal $S_D$ and $\kappa$ values along flame contour points for the highlighted flame segments. Flame contour points in Figs. 11 c,f begin at the left-most position of the flame segment and walk along the flame segment in equally spaced points. Flame wrinkles are identified in Fig. 11 c,f when curvature crosses the $\kappa = 0$ axis and indicated by the dotted vertical lines.

When analyzing each individual flame wrinkle, an inverse relationship between $S_D$ and $\kappa$ becomes apparent. In particular, as $\kappa$ decreases along concave flame wrinkles, there is a notable increase in $S_D$, while for convex flame wrinkles there is a notable decrease in $S_D$ as $\kappa$ increases. This inverse relationship is valid for both small and large changes in curvature ($|\Delta\kappa| = 0.5 - 3.0$).

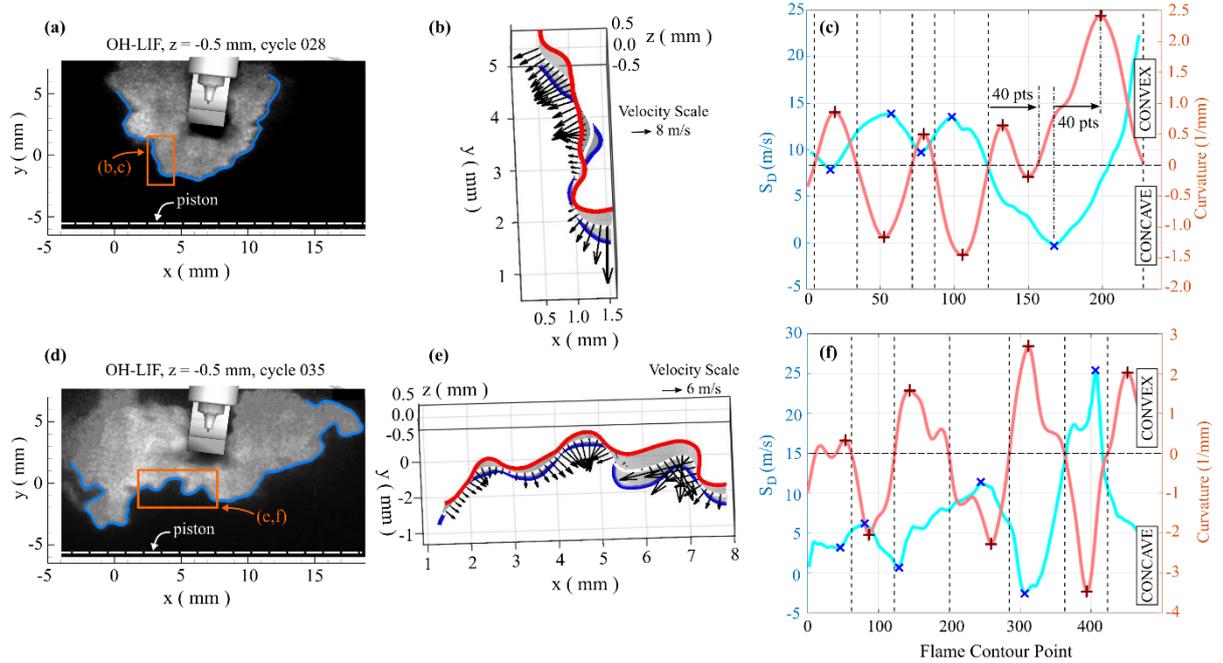

*Figure 11: $S_D$ and $\kappa$ are evaluated along selected flame segments with different degrees of flame wrinkling. (a, d) OH-LIF images (z = -0.5 mm). (b,e) $S_D$ shown along the flame contour (z = 0 mm) for the selected flame segments highlighted in a,d. Every 10th vector shown. (c,f) $S_D$ and $\kappa$ evaluated along z = 0 mm flame contours in b,e. Dotted vertical lines indicate the $\kappa = 0$ crossing. Inflection points with local maximum/minimum $S_D$ and $\kappa$ values are indicated by x and + symbols, respectively.*

In some locations, however, deviations from the inverse relationship between $S_D$ and $\kappa$ can exist. For example, cycle 28 shown in Figs. 11 a-c, initially shows that $S_D$ and $\kappa$ inflection points are nearly perfectly aligned. However, beyond flame contour point 125, the inverse relationship between $S_D$ as $\kappa$ becomes less synchronized. Along contour points 145-155, there is a brief change in $\kappa$ (convex → concave → convex). While $\Delta S_D$ becomes less negative in this region, $\Delta S_D$ does not mirror the same directional change (i.e. become positive) for this small curvature change around $\kappa = 0$. As a result, the inflection points are no longer perfectly aligned, but the inverse relationship between $\Delta S_D$ and $\Delta \kappa$ still exists. It is remarkable that the distance between the small convex → concave flame segment is the same distance that offsets the $S_D$ and $\kappa$ inflection points downstream as shown in Fig. 11c (40 flame contour points; 1 contour point = 20 μm). In this example, the small directional change in $\kappa$ across $\kappa = 0$ appears to create a delay in the response of $S_D$ with respect to changes in $\kappa$.

For cycle 35, shown in Figs. 11 d-f, the flame wrinkling is more pronounced leading to larger $\kappa$ values. While the inflection points for $S_D$ and $\kappa$ are not perfectly aligned, the inverse relationship between $S_D$ and $\kappa$ exists for flame wrinkles with $|\kappa| > 0.5$.

The relationship between $S_D$ and $\kappa$ is further evaluated for all engine cycles. In this analysis, the average derivative between the inflection point and the crossing at the $\kappa = 0$ axis is considered for each flame wrinkle as described in Eq 3.

$$\frac{d(Y)}{dx_{Flame}} = \left[\frac{Y_i-Y_1}{x_i-x_1} + \frac{Y_i-Y_2}{x_i-x_2}\right]/2 \quad (3)$$

where $Y$ indicates the parameter $S_D$ or $\kappa$, the subscript $i$ indicates values at the inflection point, and subscripts $1,2$ indicates values at the $\kappa = 0$ crossing. In this analysis, $S_D$ has units of m/s, $\kappa$ has units of 1/mm and $dx_{Flame}$ has units of mm. The derivative for each data point along the flame contour is not considered because, as shown in Fig. 11, inflection points for $S_D$ and $\kappa$ may not be perfectly aligned and will cause considerable scatter when comparing $S_D$ and $\kappa$ derivatives along each contour point of the flame.

Figure 12 a,b plots the relationship between $d\kappa/dx_{Flame}$ and $dS_D/dx_{Flame}$ for all flame wrinkles identified for 100 engine cycles at each RPM. The data is divided into four quadrants (Q1-Q4). At 800(1500) RPM, 94.1%(92.4%) of the data points lie in Q2 and Q4. Data in Q2 represents flame segments for which $S_D$ *increases* as the flame surface becomes more concave, while data in Q4 represents flame segments for which $S_D$ *decreases* as the flame surface becomes more convex. The negative correlation between $S_D$ and $\kappa$ gradients is consistent with flame theory for thermo-diffusively stable flames where the Lewis number (*Le*) is greater than unity [40-43]. In premixed combustion, *Le* is traditionally defined for the deficient species, either fuel or oxidant. In the current work, the engine was operated with stoichiometric $C_8H_{18}$-air mixtures, such that there is no clear deficient species. Bechold and Matalon [44] have suggested that both fuel and oxidant will have an effect on the effective mixture Lewis number (*Le$_{eff}$*), which is defined as:

$$Le_{eff} = 1 + \frac{(Le_E-1)+A(Le_D-1)}{1+A} \quad (4)$$

where subscripts *E,D* represent the excess and deficient species, respectively and *A* is defined as:

$$A = 1 + \beta(\varphi - 1) \quad (5)$$

where $\beta$ is the Zeldovich number and $\varphi$ is related to the equivalence ratio and is either equal to or greater than unity. The effective *Le* proposed in [44] can theoretically be used to calculate *Le$_{eff}$* for cases of stoichiometric mixtures. In such cases, $\varphi = 1$ such that $A = 1$ and *Le$_{eff}$* is effectively the weighted averaged *Le* of both substances (i.e. $Le_{eff} = (Le_{oxidant} + Le_{fuel})/2$). CHEMKIN is used in this work to calculate the thermal and mass diffusivities at the relevant engine conditions (P = 11.1 bar, T ≈ 500 K). The effective Lewis number for the $C_8H_{18}$-air mixtures in this work is *Le$_{eff}$* = 1.99. This suggests a thermo-diffusively stable flame for which the results shown in Fig. 12 are consistent with flame theory; namely $S_D$ will increase for concave flamelets and $S_D$ will decrease for convex flamelets. Until now, this inverse relationship has primarily been shown for 3D simplified chemistry- and 2D detailed chemistry-based DNS studies for much simpler flame environments [6-9, 45-48].

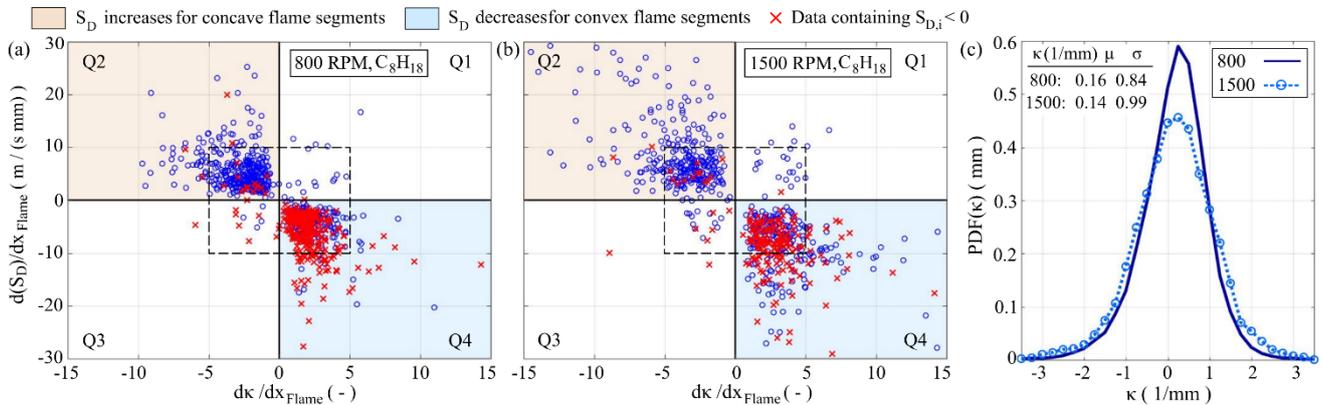

Figure 12: a,b) $d\kappa/dx$ vs. $dS_D/dx$ for individual flame segments exhibiting convex and concave flame wrinkles. Symbol **o** represents flame segments having positive $S_{D,i}$ values. Symbol **x** represents flame segments having negative $S_{D,i}$ values. c) $\kappa$ PDF for all points along flame contours. Data in Fig. 12 consists of 100 cycles at each RPM, which includes 912 and 854 flame segments at 800 and 1500 RPM, respectively.

At 800 RPM, 16.7% of the data is located outside the region $|d\kappa/dx_{Flame}| \leq 5$, $|d(S_D)/dx_{Flame}| \leq 10$ (dashed rectangle). At 1500 RPM, the data is distributed within a larger domain and 35.6% of the data fall outside of this region. The PDF of $\kappa$ along each contour point (100 cycles) is shown in Fig. 12c for each RPM. While $\kappa$ shows a slightly broader distribution towards larger $|\kappa|$ values at 1500 RPM, the analysis in Figs. 12a,b demonstrates that the $\kappa$ gradients can be significantly larger at 1500 RPM, which often yield a larger $S_D$ gradient along the flame contour. This suggest that the more severe wrinkling at the higher RPM, may directly result in higher (positive) and negative flame speeds as depicted in the PDF distributions of Fig. 6.

*4.2.1 Negative $S_D$*

It is important to emphasize that the notion of a negative $S_D$ does not pertain to rates of fuel consumption or heat release, but instead pertains to the flame displacement *relative* to the flow. Several experimental and computational studies have reported negative $S_D$ [6-9, 22, 23, 46-51]. DNS studies have identified three primary mechanisms attributed to negative $S_D$: (1) strong diffusive effects tangential to flame contours exhibiting high positive curvature [7-9], (2) high compressive or tangential strain [8, 9, 47], and (3) sensitivies of the prescribed isolevel used to identify the flame front [51].

The data points shown in Fig. 12 distinguish between flame segments that have either a positive or negative $S_D$ value at its inflection point (i.e. $S_{D,i}$ in Eq. 3). Flame segments associated with $S_{D,i} > 0$ are shown as blue circles, while flame segments associated with $S_{D,i} < 0$ are shown as red crosses. The majority of flame segments involving negative $S_{D,i}$ values are associated with convex flame curvature; at 800(1500) RPM, 89.6%(87.9%) of data with $S_{D,i} < 0$ are located in Q4. Such findings are in agreement with DNS studies [7-9, 47], which report negative $S_D$ associated with high positive flame curvature for thermo-diffusively stable flames. Gran et al [6] and Chakraborty [9] demonstrate that this phenomenon is due to strong molecular diffusive effects tangential to flame contours exhibiting high positive curvature. Chakraborty [9] demonstrates that tangential diffusive effects causing a negative $S_D$ is much stronger in the thin reaction zone regime than in the corrugated flamelet regime of premixed combustion.

Analysis of the turbulent flow field in *Sect. 3.1* and preliminary analysis of length scales using the two-point auto-correlation provide the following at the image timing employed: $l/\delta_F \sim 9 - 20$ and $u'/S_L = 6.9 - 22.2$. While these findings are limited on the Reynold decomposition methodology, these values are suggestive that the reaction takes place within the thin reaction zone regime for the imaging timing and engine conditions employed in this work. This may explain why the overwhelming majority of the data exhibiting negative $S_D$ occur for positively curved flame segments. However, we should also recognize that positive flame curvature does not directly yield a negative $S_D$; 34% and 40% of the $S_{D,i} > 0$ data are also located in Q4 for 800 and 1500 RPM, respectively.

Despite the majority of negative $S_D$ data exhibiting positive flame curvature, other factors could also contribute to negative $S_D$ values but are not evaluated in detail. For example, high tangential and compressive strain can yield a negative $S_D$, however, strain rates along the flame surface are not evaluated in this work due to the lack of gradient information in the z-direction. Since the flow field and flame propagation within the engine is highly 3D, we anticipate that z-gradients will be significant and should not be ignored. Negative $S_D$ values can also arise from measurement uncertainty, including limitations in linear approximation of the flame surface between the parallel planes. This was discussed in detail in [22] and it was shown that negative $S_D$ values still occur beyond maximum uncertainty, signifying that they are statistically significant.

### 4.3 Spatially distributed $\vec{U}_{gas}$, $S_D$ and $\vec{U}_{Flame}$

*4.3.1 Individual cycle analysis*

Thus far, the analysis has revealed the mechanisms that can lead to positive or negative, slow or fast $S_D$ and $\vec{U}_{Flame}$ velocities. In this section we evaluate how $\vec{U}_{gas}$, $S_D$ and $\vec{U}_{Flame}$ are spatially distributed throughout the FOV. In this section, $\vec{U}_{Flame}$ is reported as a positive(negative) velocity if the flame is transported to the reactants(products). This is different than previous sections where $\vec{U}_{Flame}$ was strictly positive and $\vec{U}_{Flame} \cos(\beta)$ (i.e. the flame-normal component) was evaluated as positive/negative to determine the velocity transport towards reactants/products. In this section, it is more meaningful to report the velocity magnitude rather than individual components, while still indicating the direction of the overall flame progression.

Figure 13 shows $\vec{U}_{gas}$, $S_D$ and $\vec{U}_{Flame}$ spatially distributed along flame contours for two individual cycles at each RPM. Velocity values shown along the flame contour are spatially averaged onto a 0.25 x 0.25 mm² grid. The left-most images show the flame contours superimposed onto the SPIV images. In these images three general regions (R1−R3) are highlighted to discuss local flame propagation characteristics that are common amongst recorded cycles. Region 1 is a general region downstream of the spark plug and often encompasses the tumble center. R1 often exhibits the lowest $\vec{U}_{gas}$ velocities in the FOV. Within R1, $S_D$ and $\vec{U}_{gas}$ velocities are equal in magnitude with the exception that $S_D$ is greater in some locations. This implies that both thermal diffusion and convection can equally contribute to the overall flame displacement velocity in R1.

Region 2 is a general region directly above the piston and beneath the spark plug. R2 exhibits the *sweeping* flow motion as described in *Sect. 3.1*, for which velocity magnitude increases from right-to-left. Consequently as one progresses from right-to-left along the flame contour in R2, $\vec{U}_{gas}$ exceeds $S_D$, revealing that convection becomes the more dominating mechanism for flame displacement. Both individual cycles also show local regions of $\vec{U}_{Flame} < 0$ where convection opposes the flame normal (i.e. α > 90º) and exceeds thermal diffusion transport directed towards the reactants.

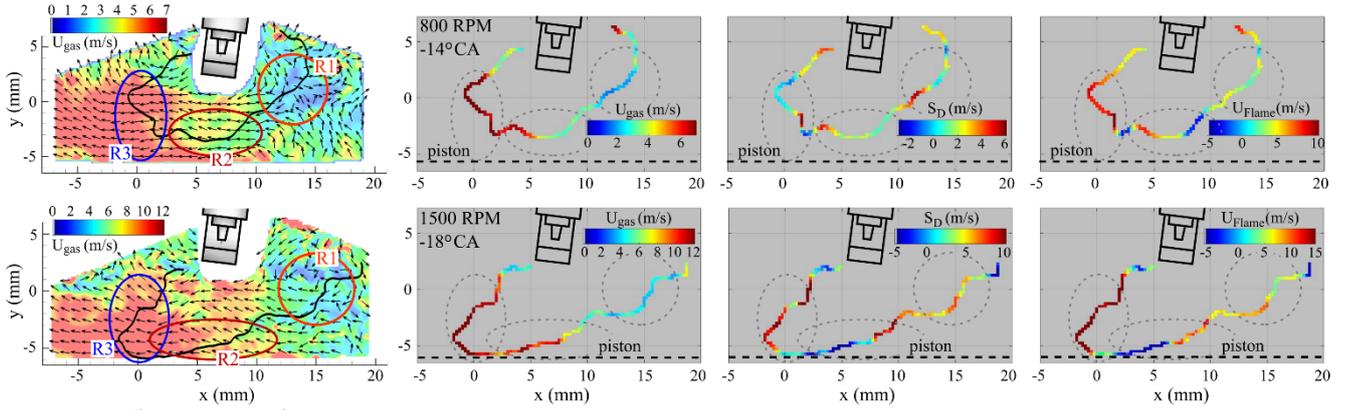

*Figure 13:* $\vec{U}_{gas}$, $S_D$ and $\vec{U}_{Flame}$ values along flame contours (z = 0 mm) for two individual cycles at 800 and 1500 RPM. Values shown along the flame contour are spatially averaged onto a 0.25 x 0.25 mm² grid. Every 6th vector shown in SPIV images.

Regions 3 is a general region containing the flame furthest to the left of the spark plug and towards the intake. The flame contour in R3 experiences the highest $\vec{U}_{gas}$ velocities in the FOV. In some locations $S_D$ is comparable to $\vec{U}_{Flame}$, but otherwise $S_D$ is primarily less than $\vec{U}_{gas}$ in R3. In most locations $\vec{U}_{gas}$ and $S_D$ both transport the flame towards the reactants such that $\vec{U}_{Flame}$ exhibits the highest (positive) values in R3. Thus, the flame typically experiences the fastest flame development in R3.

Figure 13 introduces another topic that has not been discussed until now: flame transport near solid surfaces. The individual cycle at 1500 RPM shows a portion of the flame near the piston surface. At the image timing employed approx. 10% of the images captured a flame near the piston surface. When the flame is closer than 0.4 mm from the piston surface, the convection velocity is extracted 0.1 mm away from the flame location. The flame impinges on the piston surface at x = 0.5 mm. Velocities are not evaluated at this location. To the right of the flame impingement $S_D$ and $\vec{U}_{Flame}$ are amongst the lowest values along the entire flame contour: $-4 \leq S_D \leq 2$ m/s, $-6 \leq \vec{U}_{Flame} \leq 3$ m/s. The negative $\vec{U}_{Flame}$ values are a result of $\vec{U}_{gas}$ opposing the flame normal and negative $S_D$ velocities. These negative $S_D$ velocities may indicate some aspects of flame quenching near the piston. This trend is discussed further within Fig. 14, but since these measurements were not specifically designed to study flame/wall interaction, the negative $S_D$ velocities near the piston surface are merely discussed as an observation at this stage. Improved spatial resolution of LIF and PIV such as presented in [52] or spatio-thermochemical probing of flame/wall interactions using short-pulse CARS measurements [53] are better suited to further investigate the flame behavior near solid surfaces.

*4.3.2 Ensemble-average*

Figure 14 shows the ensemble-average $\vec{U}_{gas}$, $S_D$ and $\vec{U}_{Flame}$ values spatially distributed within the FOV. Values are reported for flame contours from 100 cycles at each RPM. For an individual flame contour, velocity values are spatially averaged onto a 0.25 x 0.25 mm² grid as described for Fig. 13. Similar to *Sect.* 4.3.1, $\vec{U}_{gas}$ is reported as a magnitude (i.e. always positive), while $S_D$ and $\vec{U}_{Flame}$ are reported as positive/negative to emphasize if the flame is transported towards the reactants/products. To describe the vector orientation with respect to the flame surface, $\alpha$ and $\beta$ values are extracted along the flame contours and the ensemble-average field is shown in Fig. 15. Regions R1-R3 are highlighted in Figs. 14 and 15 to coordinate the discussion with findings already presented.

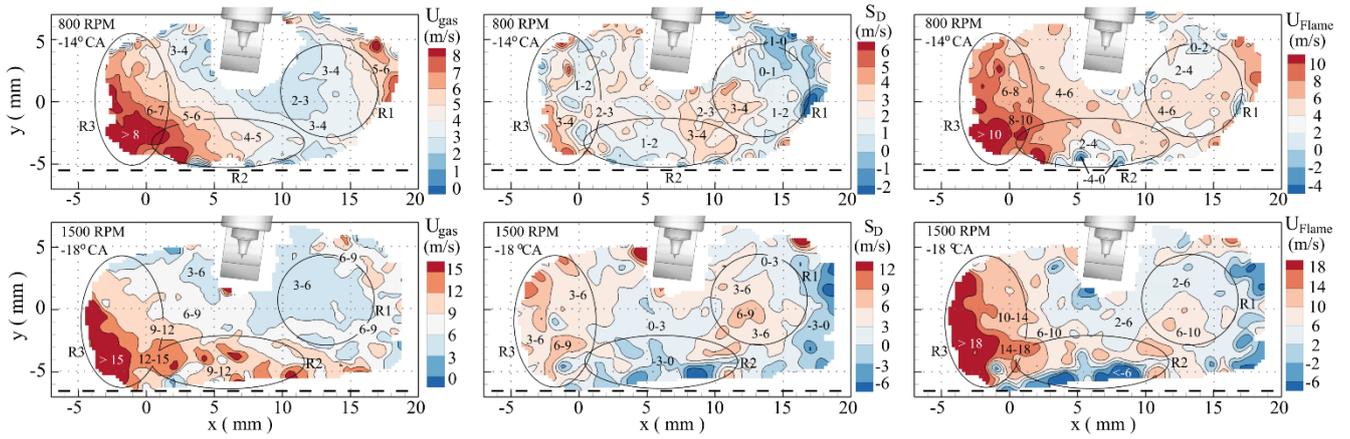

*Figure 14: Ensemble-average $\vec{U}_{gas}$, $S_D$ and $\vec{U}_{Flame}$ fields at 800 and 1500 RPM. Statistics are based on 100 cycles. Velocity values are taken along flame contours and spatially averaged onto a 0.25x0.25 mm² grid.*

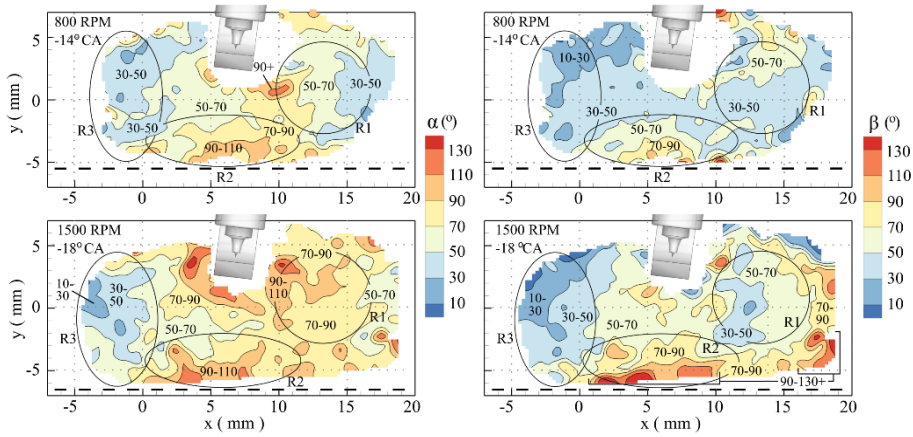

*Figure 15: Ensemble-average α and β fields at 800 and 1500 RPM. Statistics are based on 100 cycles. Values are taken along flame contours and spatially averaged onto a 0.25x0.25 mm² grid.*

The velocity maps (Fig. 14) and vector angle maps (Fig. 15) support the findings discussed in *Sect.* 4.1 and 4.3.1. At both RPMs, convection velocities are lowest to the east/southeast of the spark plug and covers a significant region of R1. Regions of high positive $S_D$ exist southeast of the spark plug and extend into R1. Within this general region, $S_D$ is equal to or slightly greater than $\vec{U}_{gas}$, indicating that both thermal diffusion and convection play an influential role in the overall flame transport. α values within R1 (and the entire FOV) are larger for 1500 RPM than at 800 RPM and signifies a more complex flame/flow interaction at higher engine speeds. Regions with α > 90° indicate regions where $\vec{U}_{gas}$ can transport the flame towards the products. Within R1 and other regions around the spark plug, pockets of α > 90° exist, but $S_D$ often exceeds $\vec{U}_{gas}$ in these regions and results in a positive flame displacement (i.e. $\vec{U}_{Flame}$ > 0).

Below the spark plug in R2, Fig. 14 shows a clear increase in $\vec{U}_{gas}$ from right-to-left with higher velocities present at 1500 RPM. In region R2, Fig. 14 reveals that convection becomes the more dominant mechanism for flame transport as $\vec{U}_{gas}$ greatly exceeds $S_D$. Additionally, Fig. 15 reveals that a majority of R2 exhibits α value between 70°–110°, which reiterates the strong *sweeping* flow motion in this region. Several locations directly above the piston exhibit α > 90°, indicating that the strong convection velocity often opposes $\vec{n}$, leading to negative flame transport. This is more pronounced at 1500 RPM. The combination of adverse convection and negative $S_D$ velocities near the piston result in the negative $\vec{U}_{Flame}$ velocities observed near the piston surface.

Flame surfaces within R3 exhibit the highest $\vec{U}_{gas}$ velocities within the FOV. While regions of high $S_D$ also exist in R3, convection is typically the more dominant transport mechanism. This is consistent for both RPMs. Moreover, Fig. 15 reveals that $\alpha$ values in R3 are amongst the lowest within the FOV. Thus, the high $\vec{U}_{gas}$ velocities are more-aligned with $\vec{n}$ such that $\vec{U}_{gas}$ and positive $S_D$ velocities both contribute towards positive flame displacement. As a result, flame displacement is shown to be the fastest within R3, particularly on the leading edge of the $\vec{U}_{Flame}$ map for both RPMs. The rapid flame development in R3 would suggest the transition from the early flame development (0-10% mass fraction burned (MFB)) to the main combustion phase (10-75% MFB) when the majority of the fuel's heat is released.

The ratio $\vec{U}_{gas}/\vec{U}_{Flame}$ and $S_D/\vec{U}_{Flame}$ are also extracted along individual flame contours. In this analysis, all velocities are considered to have positive values (i.e. purely velocity magnitude). Figure 16 shows the ensemble-average fields of these ratios. These fields help quantify the contribution of $\vec{U}_{gas}$ and $S_D$ towards the overall flame transport and appropriately highlights the differences between 800 and 1500 RPM. With an exception of a few locations, it is shown that convection often has the greatest contribution towards the overall flame transport and this contribution increases as engine speed increases. For both RPMs, convection transport is greatest in R2 and R3 where $\vec{U}_{gas}$ velocities are the highest. At 1500 RPM, convection also dominates the flame transport in the far southeast region of the FOV. Referring to Fig. 14, this region exhibits moderate-to-high $\vec{U}_{gas}$ velocities ($6 \leq \vec{U}_{gas} \leq 12$), but equally exhibits low $S_D$ velocities ($0 \leq |S_D| \leq 3$), which results in high $\vec{U}_{gas}/\vec{U}_{Flame}$ values.

For each RPM, $S_D/\vec{U}_{Flame}$ values are consistently highest directly southeast of the spark plug and within R1. In these regions, $\vec{U}_{gas}$ is the lowest and $S_D$ is the highest within the FOV. At 800 RPM, 50-60% of flame transport is governed by thermal diffusion in this region but decreases to 40-50% at 1500 RPM.

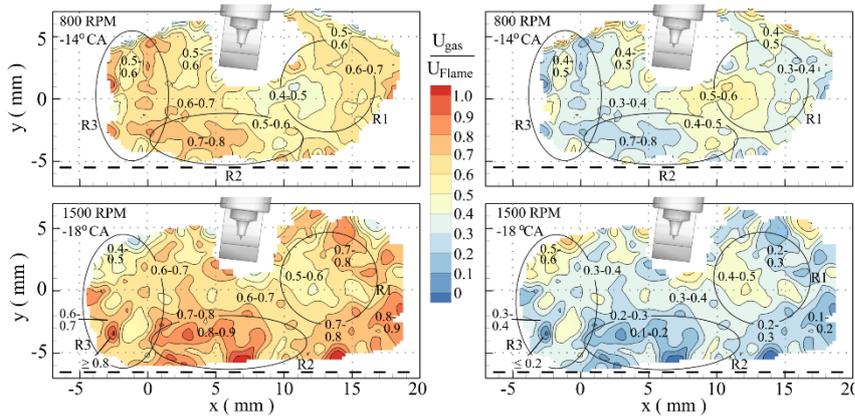

*Figure 16: Spatially distributed $\vec{U}_{gas}/\vec{U}_{Flame}$ and $S_D/\vec{U}_{Flame}$ values are used to quantify the contribution of convection and thermal diffusion towards the overall flame transport at 800 and 1500 RPM.*

*4.3.3 Statistical significance*

The velocity maps shown in Figs. 14-16 are based on 100 engine cycles. As depicted in Fig. 13, the flame contour for each cycle will only occupy a limited terrain within the FOV. Thus, not each grid point will contain velocity data from all 100 cycles. Figure 17 shows the number of data points occupied within each 0.25 x 0.25 mm² grid point to portray the statistical significance of the findings presented in Figs. 14-16.

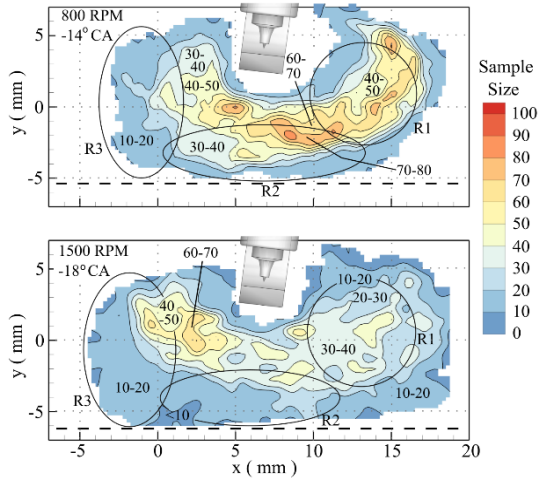

Figure 17: Number of data points evaluated within the 0.25 x 0.25 mm² grid spacing for the velocities presented in Figs. 14-15.

The number of data points is greatest within the center of the spatial domain. The maximum local sample size is larger for 800 RPM than for 1500 RPM, because the overall spatial domain is smaller at 800 RPM, while the total number of samples is similar for both RPMs (78,030 and 79,223 at 800 and 1500 RPM, respectively). Along an individual flame contour, as seen in Fig. 13, approx. 8-12 data points exist (on average) within each 0.25 x 0.25 mm² cell containing a flame contour. Dividing the local sample size by 8-12 provides a rough estimation of the number of cycles considered at a particular cell. Regions R1-R3, as discussed in Fig. 13, are highlighted within Fig. 17 to discuss the number of data samples and cycles that occur within these regions.

Grid cells with less than 10 data points only include velocity information from single cycles, however several cycles may make up the entirety of these regions. Regions with less than 10 data points are small and located along the perimeter of the spatial domain. Cells containing 10-20 data points typically include velocity information from 1−2 cycles. However, regions containing 10-20 data points per cell are large and 10's of cycles are located within such regions. For example, the region of 10-20 data within R3 contains 17 cycles for 800 RPM and 35 cycles for 1500 RPM. Similarly, the 10-20 region within R2 for 1500 RPM contains velocity data from 28 cycles. Other regions within R2 and R3 contain higher sample sizes per grid cell and comprise of even more cycles. On average, R1 contains the largest sample size per cell with up to 70-80 and 40-50 data points per cell for 800 and 1500, respectively. Overall, approx. 90% of the cycles contain a flame contour within R1 for both RPMs.

The limited number of samples within the 100 cycles is a limitation within this work and will contribute to some of the spatial variance of the distributions shown in Figs. 14-16. Although the sample size is limited, it is important to emphasize the consistent findings amongst all cycles containing flame contours in each region. Namely:

- Cycles within R1 exhibit amongst the lowest $\vec{U}_{gas}$ velocities. While thermal diffusion typically has a larger contribution towards the overall flame transport in R1, convection also plays an important role, particularly at the higher RPM.
- Flame convective transport is the most dominant within R2, which exhibits the *sweeping* flow motion above the piston surface. Within R2, $\alpha$ values often exceed 90º demonstrating that adverse convection will transport the flame towards the products. Moreover, for some cycles the flame is shown to impinge on the piston surface for which a decrease in $S_D$ is be observed. Both the convection and thermal diffusion velocities attribute to negative $\vec{U}_{Flame}$ velocities above the piston surface.

- Flame contours in R3 exhibit the largest $\vec{U}_{gas}$ and $\vec{U}_{Flame}$ velocities. Such areas exhibit rapid flame development, which may indicate the onset of the main combustion phase

While the specific location of the transport velocity behavior may be unique for this engine, its operation and image timing, the findings are presented in relation to common flow characteristics featured in most IC engines. Therefore, such findings are anticipated not to be limited to a particular engine and are appropriate to understand combustion performance and develop predictive engine models for several IC engine platforms.

## 5. Conclusions

This work presents a novel experimental dataset to study the detailed flame transport within a homogeneous charged SI engine. Dual-plane OH-LIF and SPIV were performed to resolve the thermal diffusive and convective flame transport components in a thin 3D domain. Experiments were performed at 800 and 1500 RPM during early flame kernel development when less than 5% of the fuel was consumed. Ignition and image timings were carefully selected to provide similar thermodynamic conditions at each RPM, while the difference in engine speed provided a practical environment to study flame propagation for different convection velocity and turbulence levels.

As engine speed increases, PDFs revealed that flame transport velocities show a broader distribution to both higher positive and negative velocities. Negative velocities indicate that the flame is transported towards the products rather than reactants. Analysis of flame/flow interactions was performed to describe the flame/flow configurations that lead to positive and negative flame transport. The overall flame transport is a result of the relationships between $\vec{U}_{gas}$, $S_D$ and their vector angle with the flame normal $\vec{n}$. While $S_D$ remains normal to the flame surface, the vector angle (α) between $\vec{U}_{gas}$ and $\vec{n}$ exhibits large variations, especially in regions of flame wrinkling. Positive or negative flame transport is depicted by the sign of $\vec{U}_{gas}\cos(\alpha)$ and $S_D$. When these velocities are positive, both $\vec{U}_{gas}$ and $S_D$ participate to transport the flame towards the reactants and often yield the highest $\vec{U}_{Flame}$ velocity. The measurements revealed several flame/flow relationships that describe the complex nature of flame development for turbulent flows within SI engines.

$S_D$ was evaluated with respect to flame curvature ($\kappa$) to understand the variation of $S_D$ along flame contours. The flame speed behavior is consistent with that for thermo-diffusively stable flames. Analysis also shows a strong correlation between negative $S_D$ and convex flame contours. Until now, such flame dynamics have not been studied in detail within engine environments. Moreover, at the higher engine speed, more severe wrinkling was correlated to larger changes in $S_D$, which describes the broader $S_D$ distribution towards higher positive and negative velocities.

$\vec{U}_{gas}$, $S_D$ and $\vec{U}_{Flame}$ velocities were analyzed in detail to understand their spatial distribution within the cylinder. This analysis provided the opportunity to quantify the relative contribution of convective and thermal diffusive flame transport within the FOV. The ratio $S_D/\vec{U}_{Flame}$ is greatest downstream the spark plug where unburnt gas velocities are lowest behind the spark plug and near the tumble center. Beneath the spark plug, a strong *sweeping* flow motion exists for which convection is the most-dominating mechanism of flame transport. Near the piston surface, cycles exhibit lower (and sometimes negative) $S_D$, which may indicate local flame quenching. In combination with adverse convection transport, several locations with negative $\vec{U}_{Flame}$ exist above the piston. As the flame enters into the intake domain of the cylinder, the flame/flow angle (α) is more aligned with $\vec{n}$ and the flame is rapidly transported towards the reactants by $\vec{U}_{gas}$ and $S_D$. The flame in this region exhibits rapid flame development which likely demonstrates the transition from early flame development to the main combustion phase. Overall, the convection transport becomes greater in all regions at higher RPM.

These measurements provide a first insight into the detailed 3D flame transport based on statistical analysis of instantaneous 3D time-resolved quantities within a technically relevant environment. While some local phenomena may be specific to this engine's operating conditions, most phenomena are described to common flow features (e.g. tumble, squish) in SI engines. Therefore, findings are anticipated to be a great use to the combustion and modelling community to predict more accurately the flame transport and subsequent heat release. The measurements also reveal the need to develop models that predict negative $S_d$ and $\vec{U}_{Flame}$ velocities due to flame wrinkling, adverse convection and potential flame quenching near surfaces.

**Acknowledgements**

The authors kindly acknowledge financial support from the ERC (grant 759546), DFG (SFB/Transregio 150) and EPSRC (EP/P020593/1, EP/P001661/1). A. Dreizler acknowledges financial support from the Gottfried Wilhelm Leibniz program. The authors are also especially thankful to LaVision for borrowing equipment, M.-S. Benzinger (ITT, Karlsruhe) for the 1D flamelet simulations, P. Brequigny (PRISME, Univ. Orléans) for CHEMKIN simulations, and C. Karakannas and A. Khan for their help in image processing.

**Appendix A: SPIV particle response time**

The response time of the SPIV seeding medium is evaluated to ensure that SPIV particles will appropriately follow the engine flow. The seeding particle response time is calculated as [54]:

$$\tau_p = \rho_p d_p^2 / 18\mu_{air} \quad (6)$$

where $\rho_p$ is the particle density ($\rho_{BN}$ = 2100 kg/m³), $d_p$ is the particle diameter (3 μm) and $\mu_{air}$ is the gas viscosity ($\mu_{air}$ = 2.89e-5 Ns/m² evaluated at 550K [55]). The calculated particle response time for the BN particles is $\tau_p$ = 36 μs. The seeding particles' response time should be lower than relevant timescales of the engine flow. A typical engine timescale is described by [56]:

$$\tau_e = S/V_p \quad (7)$$

where $S$ is the engine stroke (86 mm) and $V_p$ is the mean piston speed (2.3 m/s (800 RPM) or 4.3 m/s (1500 RPM)). $\tau_e$ describes the flow variation rate imposed by the time varying engine boundary conditions. Another timescale for consideration is the turbulent turn-over timescale [56]:

$$\tau_t = L/u' \quad (8)$$

where $L$ is the length scale of the size of the energy containing eddies and $u'$ is a representative RMS velocity associated with these eddies. Following the analogy of Lumley [57], $L$ can be estimated to be 1/6 the largest possible eddy (taken as the height between the piston and cylinder head; $L$ = 17.3 mm / 6). In previous studies, $u'$ is often equal to the piston velocity [56, 57]. In this study, we take $u'$ as the maximum velocity fluctuation from a Reynolds decomposition approach (see section 3). In comparison to the piston velocity, the representative $u'$ values used in this work are larger, which yields shorter $\tau_t$ timescales.

Considering these timescales, the Stokes number is calculated:

$$St = \tau_p / \tau_{e,t} \quad (9)$$

As can be seen in Table 2, Stokes numbers range from 0.04 – 0.1, with the highest values associated with 1500 RPM due to larger $u'$ values. The maximum Stokes number being smaller than unity demonstrates that the SPIV particles will accurately follow the engine flow in this study.

*Table 2: Particle time scales, flow time scales and representative Stokes numbers.*

|          | $\tau_p$ (μs) | $\tau_e$ (μs) | $\tau_t$ (μs) | $St_e$   | $St_t$   |
|----------|---------------|---------------|---------------|----------|----------|
| 800 RPM  | 36            | 37,400        | 580           | 9.6e-04  | 6.2e-02  |
| 1500 RPM | 36            | 20,000        | 320           | 1.8e-03  | 1.1e-01  |